\documentclass[structabstract]{aa}  
\usepackage{graphicx}
\usepackage{natbib}
\bibpunct{(}{)}{;}{a}{}{,}
\usepackage{txfonts}
\begin{document}
   \title{Low frequency observations of the radio nebula produced by the giant flare from SGR 1806-20}

   \subtitle{Polarimetry and total intensity measurements}

   \author{H. Spreeuw, B. Scheers \& R.A.M.J. Wijers}
   \institute{Astronomical Institute "Anton Pannekoek", University of Amsterdam,
              Kruislaan 403, 1098 SJ  Amsterdam, The Netherlands\\
              \email{[j.n.spreeuw,l.h.a.scheers,r.a.m.j.wijers]@uva.nl}\\
             \\
}

   \date{Received September 15, 1996; accepted March 16, 1997}

 
  \abstract
   {The 2004 December 27 giant flare from SGR 1806-20 produced a radio nebula that was detectable for weeks. It was observed at a wide range of radio frequencies. }
   {To investigate the polarized signal from the radio nebula at low frequencies and to perform precise total intensity measurements.} 
   {We made a total of 19 WSRT observations. Most of these were performed quasi simultaneously at either two or three frequencies, starting 2005 January 4 and ending 2005 January 29. We reobserved the field in 2005 April/May, which facilitated an accurate subtraction of background sources.}
   {At 350 MHz, we find that the total intensity of the source is lower than expected from the GMRT 240 MHz and 610 MHz measurements and inconsistent with spectral indices published previously. Our 850 MHz flux densities, however, are consistent with earlier results. There is no compelling evidence for significant depolarization at any frequency. We do, however, find that polarization angles differ substantially from those at higher frequencies. }
   {Low frequency polarimetry and total intensity measurements provide a number of clues with regard to substructure in the radio nebula associated with SGR 1806-20. In general, for a more complete understanding of similar events, low frequency observations can provide new insights into the physics of the radio source.}

   \authorrunning{H. Spreeuw et al.}
   \titlerunning{SGR 1806-20 radio flux measurements}
   \maketitle
%

\section{Introduction}

  The 2004 December 27 flare from the Soft-$\gamma$-ray Repeater SGR 1806-20 was a major event in astronomy in a number of ways. First of all by the energy of the explosion: the brightest flash of radiation from beyond our solar system ever recorded. This is how it caught the attention of a larger audience. Secondly, because the flare provided new observational data about a known class of objects: magnetars, i.e., strongly magnetized neutron stars \citep[see, e.g.,][]{Hurley2005}. Also, it led to speculation about a possible link with $\gamma$-ray bursts (GRBs) \citep[see, e.g.,][]{Tanvir2005}. Theorists investigated the connection between the magnetic field and the explosion \citep[see, e.g.][]{Blandford2005}. Other focused on modeling the fireball and the afterglow \citep[see, e.g.,][]{Nakar2005,Dai2005,Wang2005}. Astronomers performed a number of follow-up observations at various wavelengths \citep{Rea2005,Israel2005,Palmer2005,Schwartz2005,Fender2006}. In particular, the flux from the radio nebula produced by the explosion \citep{Gaensler2005a,Cameron2005, Taylor2005} was measured very frequently in 2005 January. These observations focused on total intensity measurements at various radio wavelengths and on polarimetry at 8.5 GHz. Some polarimetry was done at lower frequencies, but without the proper correction for the leakages \citep{Gaensler2005b}. We have performed accurate polarimetry at 350, 850 and 1300 MHz. Also, we were able to measure the Stokes I flux from the radio nebula at 350 and 850 MHz more precisely by observing the same field again in 2005 April/May. In this way, we could properly subtract the background sources from the (u, v) data of the 2005 January observations. We compare our measurements with those at nearby frequencies.


\begin{table}
\caption{Summary of these 19 WSRT observations}
\centering
\begin{tabular}{c c c c}
\hline \hline
Epoch      & Days since & Time on     & Frequency \\
           & burst      &  source (min)& (MHz)    \\
\hline
January 4  &  7.6  &   54   & 1300 \\
January 4  &  7.6  &   91   & 350  \\   
January 5  &  8.6  &   462  & 850 \\      
January 7  &  10.5 &   107  & 1300  \\
January 7  &  10.5 &   107  & 350  \\
January 7  &  10.5 &   107  & 850  \\
January 10 &  13.6 &   78   & 1300 \\
January 10 &  13.6 &   71   & 350  \\
January 10 &  13.6 &   71   & 850  \\
January 16 &  19.6 &  181   & 350 \\
January 16 &  19.6 &  181   & 850 \\
January 20 &  23.6 &  181   & 350 \\
January 20 &  23.6 &  165   & 850 \\ 
January 23 &  26.6 &  198   & 350 \\
January 23 &  26.6 &  198   & 850 \\
January 29 &  32.6 &  196   & 350 \\
January 29 &  32.6 &  196   & 850 \\
April 30/May 1 & 124.3 & 444 & 350 \\
May 2          & 125.2 & 464 & 850 \\
\hline
\label{tab:Obs}
\end{tabular}
\end{table}

\section{Observation and data reduction}
\subsection{General}
A total of 19 observations were performed in January, April and May of 2005. Four of these, on January 16, 20, 23 and 29 were alternating between 350 and 850 MHz. On January 7 and 10 scans at 1300 MHz were also included. On January 4 we observed at 350, 650 and 1300 MHz, but the 650 MHz data was not used. A summary is shown in table \ref{tab:Obs}. \\
We used AIPS \citep{Greisen2003} and ParselTongue \citep{Kettenis} scripts for the reduction of all 19 datasets. The Westerbork Synthesis Radio Telescope (WSRT) was used for all observations. The WSRT is a linear array with 14 equatorially mounted 25-m dishes equipped with linear feeds. Its maximum baseline is 2.7 km.  All datasets recorded four polarization products with 8 IFs. 3C286 was observed before the target and 3C48 after. RFI was excised from the spectral line data using the AIPS task 'SPFLG'.\\ Calibration was done in four steps. First we determined the variation in system temperature as a function of time (and therefore also as a function of position on the sky),
using the intermittent firing of a stable noise source. Next we performed a bandpass calibration using the AIPS task 'BPASS' using either 3C48 or 3C286 or both. We applied the bandpass solution using the AIPS task 'SPLAT'. After that, we performed an external absolute gain calibration using an assumed flux for 3C48 by running the AIPS tasks 'SETJY' and 'CALIB'. 'SETJY' was set to use the
absolute flux density calibration determined by \citet{Baars} and the latest (epoch 1999.2) polynomial coefficients for interpolating over frequency as determined at the VLA by NRAO staff. Finally, we self-calibrated the data for time variations in the relative complex gain phase and amplitude. \\
Polarization calibration was performed by running the AIPS task 'LPCAL' on 3C48 and 'CLCOR' to correct for the instrumental XY phase offset. Generally, we followed the scheme for data reduction of WSRT data in AIPS as outlined by Robert Braun \footnote{See http://www.astron.nl/radio-observatory/astronomers/analysis-wsrt-data/analysis-wsrt-dzb-data-classic-aips/analysis-wsrt-d}, although we ran some AIPS tasks differently depending on frequency. Those differences mainly involved the details of polarization calibration. For instance, the leakage terms ("D terms") of the WSRT IFs are channel dependent, as pointed out by \citet[][paragraph 3.2]{Brentjens2008}. We took account of this, by first averaging groups of 5 channels through the AIPS task 'SPLAT'. Next, we ran 'UVCOP' to make separate datasets from the averaged channels.  After that, we ran 'LPCAL' and 'CLCOR' on each of these separately before applying the feed and XY instrumental phase offset corrections by again running 'SPLAT'. \\
Before imaging Stokes Q and Stokes U and before merging the datasets from 5 channel averaging back together through 'DBCON', we applied a ParselTongue script for "derotation" to the residual data, i.e., the (u, v) data where all sources except the target were removed, by running the AIPS task 'UVSUB'. The original Aips++ glish script was kindly given to us by G. Bernardi; we modified and translated it to a Python/ParselTongue script on a channel by channel basis. The "derotation" of the visibilities is absolutely necessary, since the rotation measure (RM) of SGR 1806-20 is large, 272 $\mathrm{rad}/\mathrm{m}^2$ \citep{Gaensler2005a}. This means that the polarized signal would vanish if all IFs were imaged simultaneously. For the 350 MHz data, one really needs the derotation of the visibilities to be performed on a channel per channel basis, because the imaging of even one single IF would result in a severly corrupted measurement and underestimate of the fractional linear polarization.  The uncertainty in this RM \citep[10 $ \mathrm{rad}/\mathrm{m}^2$, see][]{Gaensler2005a} is too large for accurate polarization angle measurements, especially at frequencies below 1 GHz. For this reason we determined the RM more accurately, by fitting the $\sin{2\cdot \mathrm{RM} \cdot\lambda^2}$  spectrum of either Stokes U or Stokes Q to its measured values at the 8 wavelengths $\lambda$ corresponding to the IFs near 350 MHz and 850 MHz. The contribution to this RM from the ionosphere is naturally included in this fit, at least the part that did not vary during the observation run. We checked the output of the AIPS task 'TECOR' for any significant variations in the ionospheric Faraday rotation during every observing run. The ionospheric Faraday rotation computed by 'TECOR' is considered accurate since it does not use a model for the ionosphere but actual data from the CDDIS archive. We did not apply the ionospheric corrections from 'TECOR' to our data because it implicitly assumes that one has recorded data from circular feeds.\\
It should be clear from table \ref{tab:Obs} that the maximum observing time is 7.7 h due to the low declination of the source. Hence, the (u,v) coverage is sparse for all observations, since linear arrays like the WSRT ideally have 12h runs. The worst coverage was at three epochs when we alternated between three frequencies.

\subsection{Detailed desciption of the datasets}
\subsubsection{Observations at 350 MHz}
The 350 MHz observations were performed on January 4, 7, 10, 16, 20, 23 and 29 and April 30/May 1 of 2005. The last observation was made to make an accurate subtraction of background sources possible. This mainly concerns the subtraction of the Luminous Blue Variable discussed in Supplementary Table 1 of \citet{Gaensler2005a}. The time resolution of all observations, except the first and the last was 30s. On January 4 and April 30/May 1 the sampling times of the visibilities was 60s.  The bandwidth per IF was 10 MHz, separated 8.75 MHz from each other and centered on frequencies of 315.00, 323.75, 332.50, 341.25, 350.00, 358.75, 367.50 and 376.25 MHz. The IFs were split into 64 channels, each 156.25 kHz wide, except for the April 30/May 1 observation. For that observation, the IFs were split into 128 channels, each 78.125 kHz wide. We used an automated flagger for the initial editing of our data: WSRT flagger\footnote{http://www.astron.nl/~renting/}. 3C286 was included in the external gain calibration, along with 3C48. This was trivial, since 3C286 is unpolarized at this frequency. The assumed fluxes for 3C48 and 3C286 in the lowest frequency IF were 43.889 and 26.106 Jy, respectively. \\
The April 30/May 1 observation has the best (u,v) coverage. After performing 10 iterations of self calibration on this dataset the rms noise in the final image was $2.5 \mathrm{mJy/beam}$. Its clean components were used to solve for the gain phases and amplitudes of the other datasets using a rather sophisticated scheme. First, a deconvolution of each of the 2005 January datasets was done in order to subtract the central region containing the radio nebula and the LBV, using the AIPS tasks 'IMAGR', 'CCEDT' and 'UVSUB'. The residual data were calibrated on the April 30/May 1 model which had the clean components from the central region removed. The gain phase and amplitude solutions were then copied and applied to the original 2005 January datasets. It this way we made sure that the Stokes I flux from SGR 1806-20 would not be reduced by calibrating on a model from an observation months after the flare. As explained in section \ref{1300descr}, amplitude self calibration could also reduce the Stokes Q flux. However, due to the large RM of the source and because we use 45 of the available 64 channels, the Stokes Q flux almost completely vanishes in a single IF at 350 MHz. Thus this problem does not occur, at least not before "derotation".\\
PSR 1937+21 was observed in between SGR 1806-20 and 3C48 for polarization calibration. This polarization calibration technique is decribed in detail by \citet[][paragraph 3.2]{Brentjens2008}. Since the RM of this pulsar is positive, Stokes Q should be $90 \degr$ ahead of Stokes U with increasing $\lambda^2$, as noted by \citet[][paragraph 2.3]{Brown2009}.\\

\subsubsection{Observations at 850 MHz ("UHF high")}
We observed SGR 1806-20 on January 5, 7, 10, 16, 20, 23, 29 and May 5 of 2005. The last observation was performed to make an accurate subtraction of background sources possible. The time resolution of all observations, except for the first and the last, was 30s. The sampling time of the visibilities on January 5 and May 5 was 60s.
The bandwidth of the eight IFs is 10 MHz, they were separated exactly 10 MHz from each other and ranging from 805 to 875 MHz. Each IF was split in 64 channels with a width of 156.25 kHz, except for the May 1/2 data that were split into 128 channels of 78.125 kHz. The external gain calibration was performed using an assumed flux for 3C48 of 24.240 Jy for the lowest frequency IF. 
The 850 MHz were reduced in almost the same way as the 1300 MHz data. Only polarization calibration was performed slightly differently. Since the Stokes Q (and U) of 3C286 are not known for the "UHF high" frequencies, when the task 'CALIB' was run on this calibrator, it was set to solve for gain phases only and not for gain amplitudes.

\subsubsection{Observations at 1300 MHz}
\label{1300descr}
We observed SGR 1806-20 at 1300 MHz on January 4,7 and 10 of 2005. The total intensity measurements have already been published \citep[see][]{Gaensler2005a}, so we focused on the polarized signal. However, we did check that our Stokes I fluxes agreed with those previously published. \\
On 2005 January 4 visibilities were recorded every 60s, on January 7 and 10 every 30s.
The eight 20-MHz IFs were centered on frequencies of 1255, 1272, 1289, 1306, 1323, 1340 and 1357 MHz. Each IF was split in 64 channels with a width of 312.5 kHz. The external gain calibration was performed using an assumed flux for 3C48 of 17.388 Jy for the lowest frequency IF. 
3C286 was also included in the external gain amplitude and phase calibration using an assumed flux of 15.550 at 1255 MHz. 3C286 is linearly polarized. We took account of this and of the usual "AIPS for linear feeds" projection (R$\rightarrow$X,L$\rightarrow$Y) by placing the assumed Stokes Q flux of 3C286 (0.594 Jy  at the lowest frequency IF) with a minus sign at the position of Stokes V in the AIPS SU table. For the other IFs we kept the same ratio between Stokes I and Stokes Q. In this way we could use 3C286 not only for fixing the instrumental XY phase offset, but also for external gain calibration. \\
Self-calibration was run to solve for the gain phases only, since solving for the amplitudes could reduce the Stokes Q flux. The AIPS task 'CALIB' cannot be set to run simultaneously on a Stokes I and Stokes Q model. Obviously, when 'CALIB' is run on a Stokes I model, it implicitly assumes that Q=0. Consequently, the same model is used to derive the X gains from the XX visibilities as the Y gains from the YY visibilities, while in fact XX=I-Q and YY=I+Q, so different models should be used. When solving for gain phases only, the error made is generally considered acceptable.

\begin{table}
\caption{Stokes I flux measurements at 350 and 850 MHz; clean components from the 2005 April 30/May 1 and May 1/2 observations were subtracted}
\label{tab:350MHzStokesI}
\centering
\begin{tabular}{| c | c | c | c | c | c |}
\hline 
 &  & \multicolumn{2}{c|}{{\bf 350 MHz}}                       & \multicolumn{2}{c|}{{\bf 850 MHz}} \\
\cline{3-6}
Epoch      & Days     & Stokes I            & 1 $\sigma$          & Stokes I            & 1 $\sigma$ \\
(2005      &  since    & flux dens.        & error               & flux dens.        & error      \\
date)           & burst          & $\mathrm{mJy/beam}$ & $\mathrm{mJy/beam}$ & $\mathrm{mJy/beam}$ & $\mathrm{mJy/beam}$ \\
\hline \hline
Jan. 4  &  7.6  &  186  & 20  &     &     \\
Jan. 5  &  8.6  &       &     & 157 & 10  \\
Jan. 7  &  10.5 &  84   & 10  & 97  & 28  \\
Jan. 10 &  13.6 &  78   & 10  & 50  & 22  \\
Jan. 16 &  19.6 &  16   & 10  & 35  & 14 \\
Jan. 20 &  23.6 &  7    & 10  & 21  & 8  \\
Jan. 23 &  26.6 &  13   & 10  & 17  & 7  \\
Jan. 29 &  32.6 &  1    & 10  & 22  & 9 \\
\hline 
\end{tabular}
\end{table}

\begin{figure*} 
\centering
     \includegraphics{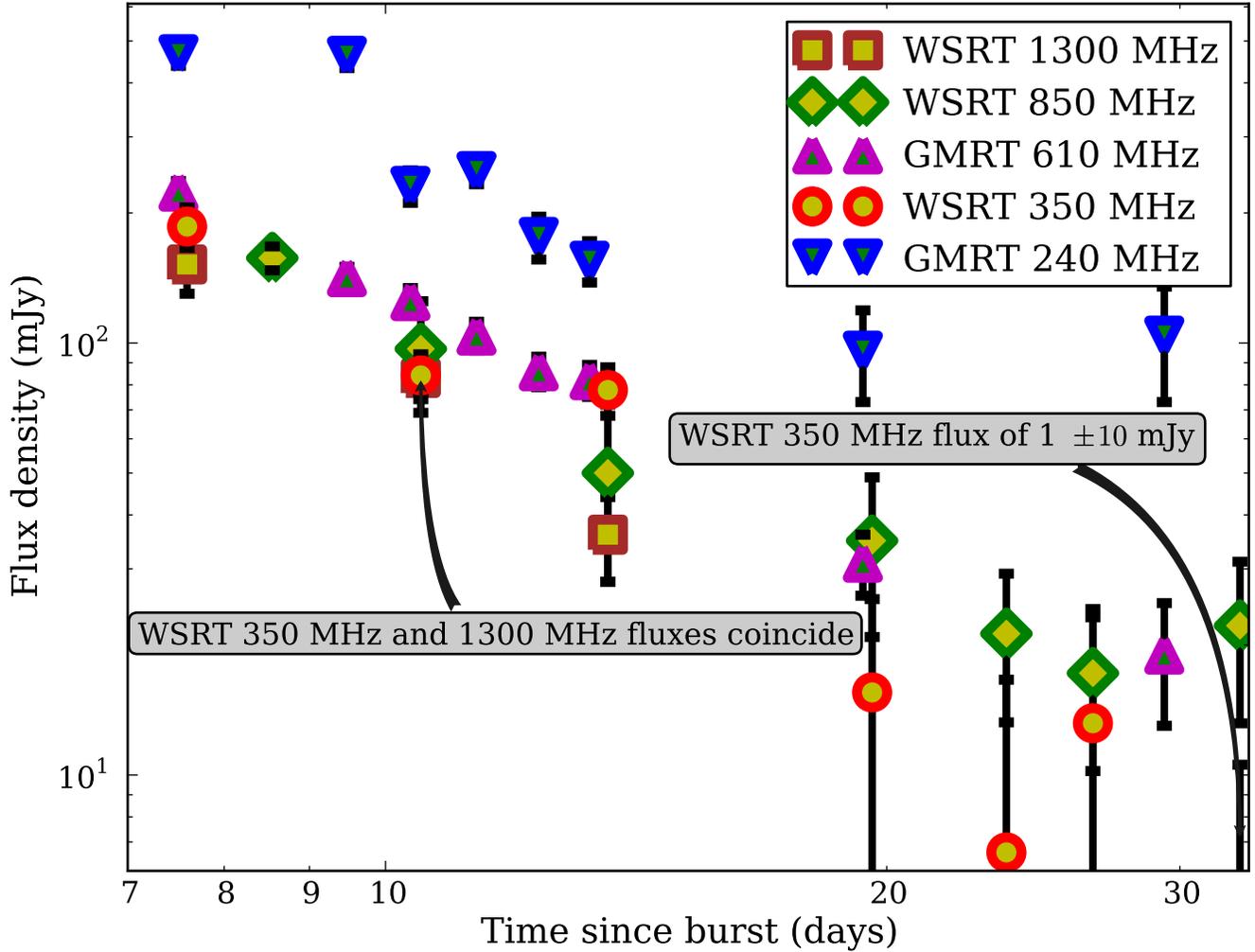}
     \caption{Comparison between the 240, 350, 610 and 1300 MHz fluxes of the radio nebula associated with SGR 1806-20. The 1300 MHz fluxes were published previously \citep{Gaensler2005a}.}
     \label{fig:fluxPband}
\end{figure*}

\section{Results}
\subsection{Total intensity measurements}
The total intensity flux measurements at 350 MHz were done by fitting a Gaussian of the same shape and size as the restoring beam to the (fixed) location of SGR 1806-20 in the Stokes I images. This was done by the AIPS task 'IMFIT'. We used the position from \citet{Gaensler2005a} ($\mathrm{\alpha=18h08m39.343s, \delta=-20\degr24\arcmin39.8\arcsec}$) for the fits. The results are summarized in table \ref{tab:350MHzStokesI}. The error bars are conservative estimates from measurements of the residuals of bright sources in the field. The actual rms noise in these images is much lower, around $3 \mathrm{mJy/beam}$, which is about the same as the error from 'IMFIT'.\\

\begin{figure} 
\centering
     \resizebox{\hsize}{!}{\includegraphics{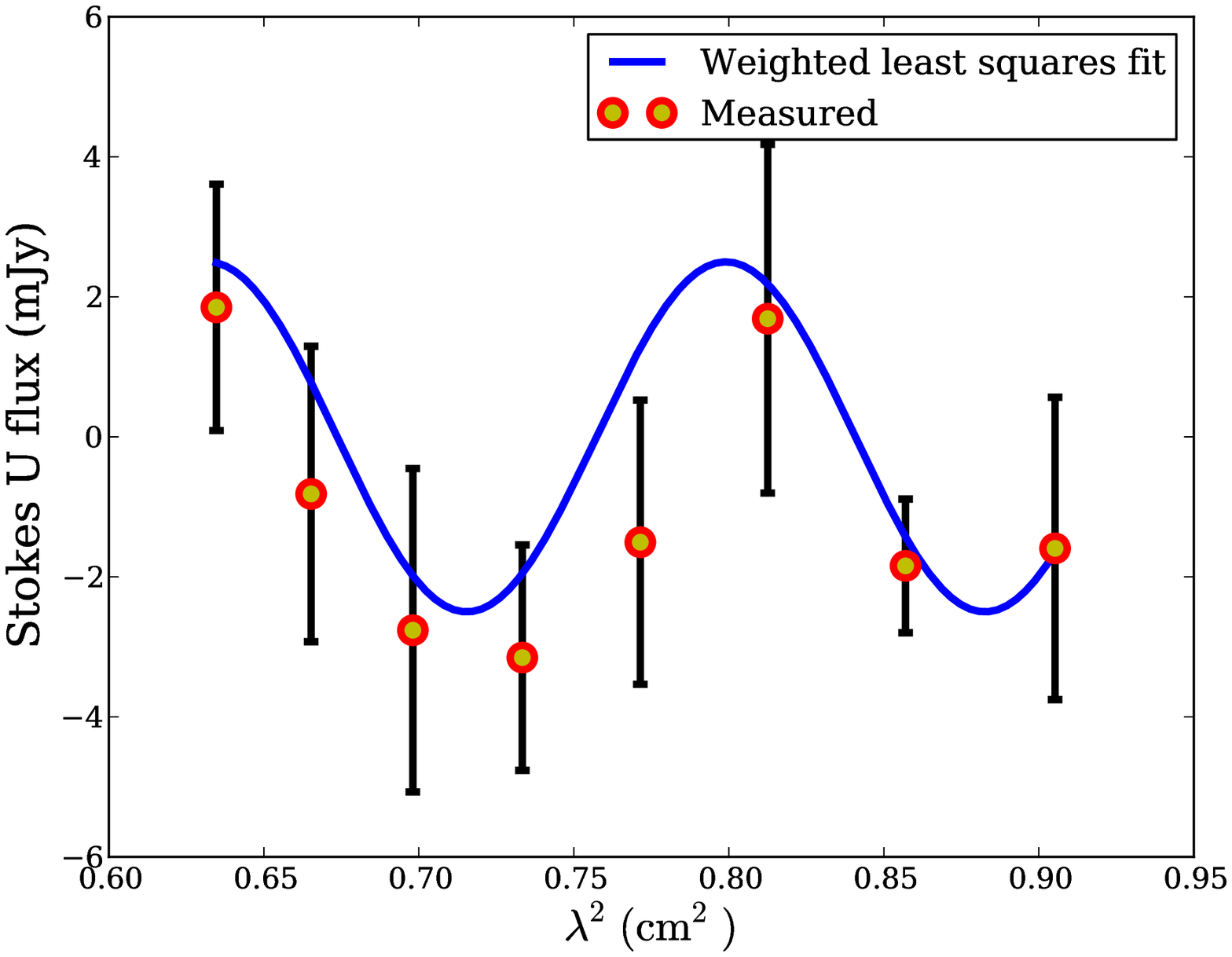}}
     \caption{Determining the rotation measure of SGR 1806-20 by fitting the sinusoidal Stokes U spectrum. Here, the fit was made to the values of Stokes U on 2005 January 04 at the wavelengths corresponding to the 8 IFs near 350 MHz after the visibilities were "derotated" by an angle corresponding to an RM of $-272\pm10$ $\mathrm{rad/m^2}$. We used gnuplot to fit the function $A\cdot \sin(2\cdot \mathrm{RM} \cdot \lambda^2+\theta)$ for three free parameters $A, \mathrm{RM}$ and $\theta$. The correction to the RM from this fit is $-18.76 \pm 1.82$ $\mathrm{rad/m^2}$. The reported reduced $\chi^2$ is 0.69.}
     \label{fig:04janfit350MHz}
\end{figure}

\begin{table}
\caption{RM measurements of SGR1806-20.}
\label{tab:RMmeasurements}
\centering
\begin{tabular}{| c | c | c | c | c |}
\hline 
Epoch      & Days     & Frequency  & Measured             & 1 $\sigma$          \\
(2005      &  since   & (MHz)      & RM                   & error                \\
date)      & burst    &            & ($\mathrm{rad/m^2}$) & ($\mathrm{rad/m^2}$) \\
\hline \hline
Jan. 4     &  7.6     &  350       & 253.24  & 1.82  \\
Jan. 5     &  8.6     &  850       & 253.14  & 12.43 \\
Jan. 7     &  10.5    &  350       & 253.65  & 1.05  \\
Jan. 10    &  13.6    &  350       & 261.74  & 2.04  \\
\hline 
\end{tabular}
\end{table}

\begin{table*}
\caption{Polarimetric measurements of SGR 1806-20}
\centering
\begin{tabular}{c c c c c c c c c c}
\hline \hline
Epoch      & Days since & Frequency & $\sqrt{Q^2+U^2}$  & 1 $\sigma$ & Polarization & 1 $\sigma$ & Polarization & 1 $\sigma$ & Reduced\\
(2005)     & burst      & (MHz)       &  ($\mathrm{mJy/beam}$) & error ($\mathrm{mJy/beam}$) & fraction ($\%$)& error ($\%$)& angle ($\degr$) & error ($\degr$) & $\chi^2$  \\
\hline
Jan. 4  &  7.6  &   350 & 2.68 & 0.81 & 1.44 & 0.46 & 103& 39 & 0.70 \\
Jan. 4  &  7.6  &  1300 & 0.71 & 0.66 & 0.47 & 0.44 & 31 & 26 & 1.57 $^{\mathrm{1}}$\\
Jan. 5  &  8.6  &   850 & 2.22 & 0.28 & 1.41 & 0.20 & 96 & 7  & 0.85 \\
Jan. 7  & 10.5  &   350 & 2.30 & 0.64 & 2.73 & 0.82 & 69 & 38 & 0.33 \\
Jan. 7  & 10.5  &   850 & 1.71 & 0.52 & 1.76 & 0.74 & 44 & 9  & 0.65 \\
Jan. 7  & 10.5  &  1300 & 1.90 & 0.37 & 2.29 & 0.48 & 50 & 7  & 1.06  \\
Jan. 10 &  13.6 &   350 & 1.14 & 0.87 & 1.46 & 1.13 & 36 & 47 & 0.79 \\ 
Jan. 10 &  13.6 &  1300 & 1.31 & 0.40 & 3.65 & 1.38 & 69 & 12 & 1.28 \\
\hline
\label{tab:linpolfrac}
\end{tabular}
\begin{list}{}{}
\item[$^{\mathrm{1}}$] Poor fit.
\end{list}
\end{table*}

\begin{figure} 
\centering
     \resizebox{\hsize}{!}{\includegraphics{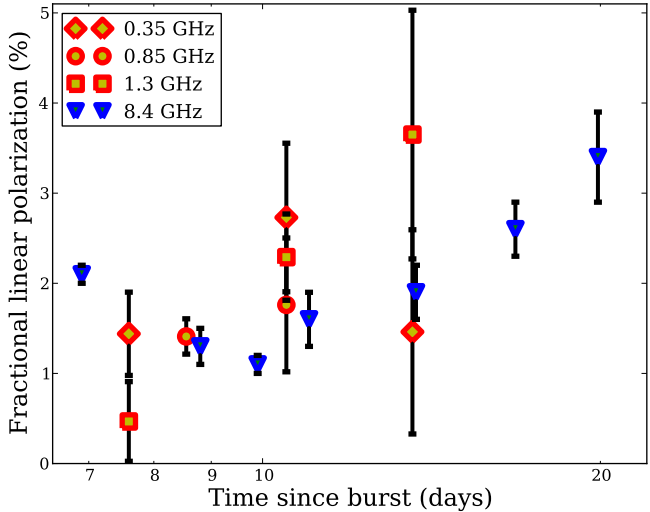}}
     \caption{Comparison between linear polarization fractions at 350, 850, 1300 and 8400 MHz}
     \label{fig:linpolfracs}
\end{figure}

\begin{figure} 
\centering
     \resizebox{\hsize}{!}{\includegraphics{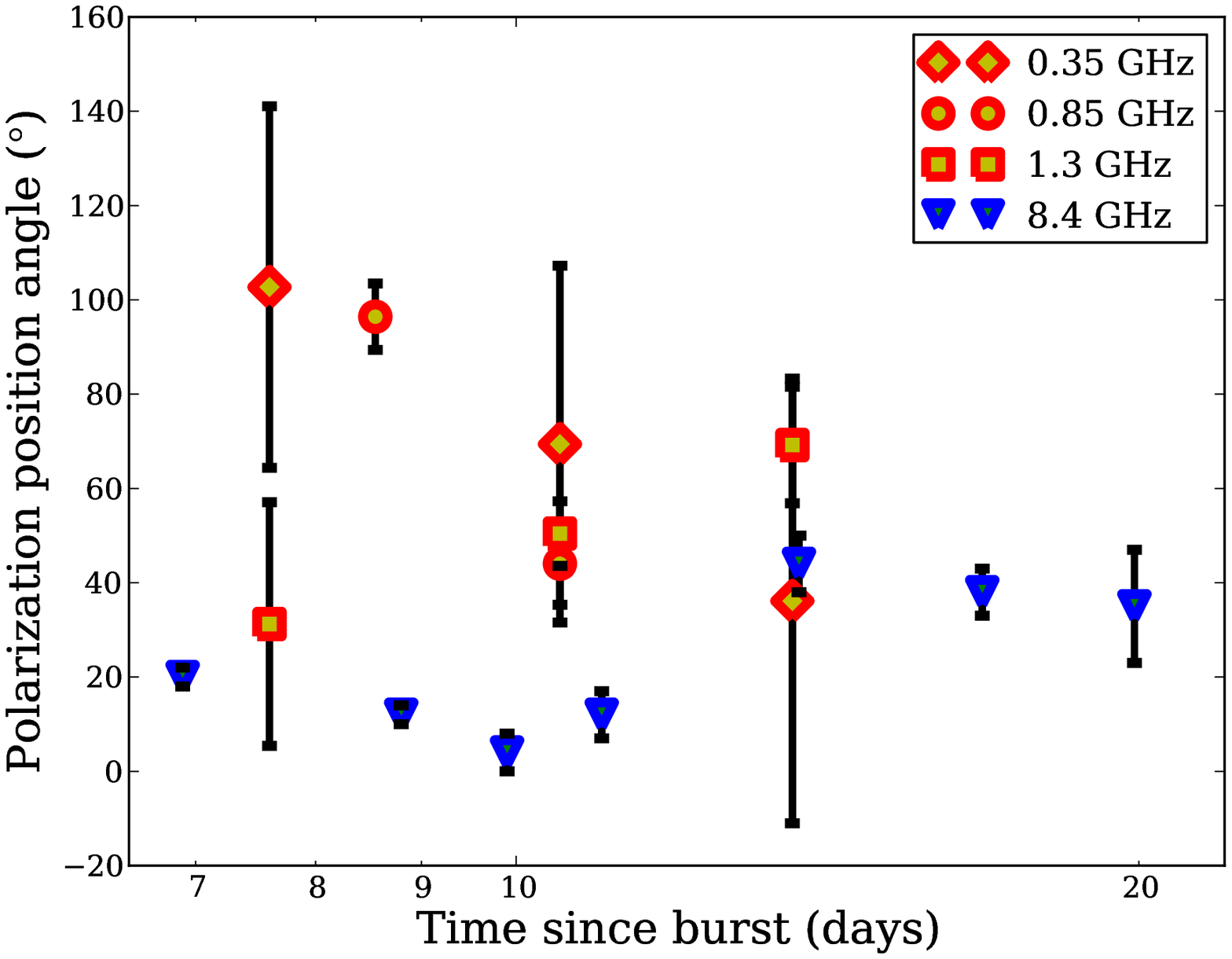}}
     \caption{Comparison between the polarization angles at 350, 850, 1300 and 8400 MHz}
     \label{fig:polangles}
\end{figure}

\subsection{Polarimetry}
\subsubsection{General}
\label{par:polgeneral}
Polarimetry was performed on 2005 January 4, 5, 7 and 10. Although all of our observations recorded full Stokes, we anticipated that it would not be possible to detect the polarized signal from SGR 1806-20 on later dates, since the total intensity drops rapidly. Also, we did not expect polarization fractions to exceed the values given by \citet[][table 2]{Taylor2005}. 

\subsubsection{Determining the RM of SGR 1806-20}
As noted before, the rotation measure (RM) as measured by \citep[][$272\pm10$ $\mathrm{rad/m^2}$]{Gaensler2005a} has a rather large error bar which translates into a polarization angle uncertainty at 1300 MHz of $30.5\degr$. At 850 MHz this is even $71.4\degr$. Naturally, the RM should be determined more accurately before polarization angles are to be measured.\\
This can be done by plotting Stokes U or Q fluxes of SGR 1806-20 as a function of frequency and fit for the RM. We are in the advantageous position that these WSRT observations were performed with eight IFs. Over a wide span of frequencies there are many turns of Stokes U (or Q) since its spectrum is sinusoidal as a function of $\lambda^2$. This effect is largest at low frequencies: at 1300 MHz, there is less than one cycle of $A\cdot \sin(2 \cdot \mathrm{RM} \cdot\lambda^2+\phi)$, at 850 MHz there are almost two cycles and at 350 MHz there are 23 cycles. It is evident that the most accurate measurement can be made at the lowest frequency, if there is sufficient signal to noise. Fortunately, we could detect polarized signal at 350 MHz from all three observations on 2005 January 4,7 and 10 after an initial "derotation" of our visibilities using the RM from \citet[][272 $\mathrm{rad/m^2}$]{Gaensler2005a}. This initial derotation prevents diminution of the polarized signal in a single IF. At 850 MHz this initial derotation was not necessary. The noise levels at that frequency were such that detecting a polarized signal was only possible on 2005 January 5 and 7, but the latter observation yielded a very poor constraint on the RM, so we left it out. The 1300 MHz data also gave very poor constraints on the RM, thus in determining the weighted mean RM we ignored those, too. For the other observations, we plotted Stokes U per IF and solved for the RM (850 MHz) or the correction to the RM (350 MHz), as illustrated in figure \ref{fig:04janfit350MHz}. The results are shown in table \ref{tab:RMmeasurements}. It turned out that the noise levels in all of the Stokes Q maps were much higher than in the Stokes U maps, so we did not use them. In determining the weighted mean RM we also took into account the measurement by \citet[][$272\pm10$ $\mathrm{rad/m^2}$]{Gaensler2005a}  From this set of five measurements we derived an RM of $255.01 \pm 0.83$ $\mathrm{rad/m^2}$.  It should be clear that, with regard to the 350 MHz RM measurements, the fits give the same reduced $\chi^2$ for both the positive and the negative correction to the initial "derotation". We removed those ambiguities by considering the Stokes U measurements near 850 MHz data on 2005 January 5. The fit to that data gave an RM of $253.14\pm12.43$ $\mathrm{rad/m^2}$ which made all of the positive RM solutions to the 350 MHz data very unlikely ($\simeq 3.0 \sigma$ level for January 4 and 7). \\
It is evident that the contribution of the ionosphere to the RM, $\mathrm{RM_{ion}}$, is included in all fits. For the 2005 January 4, 5, 7 and 10 observations, $\mathrm{RM_{ion}}$ as reported by the AIPS task 'TECOR', is the range $2.1\pm 0.4$ $\mathrm{rad/m^2}$. Consequently, the interstellar RM is given by $\mathrm{RM_{int}}=255.01-2.1=252.91 \pm 0.92$ $\mathrm{rad/m^2}$.  \\

\subsubsection{Polarization fractions and position angles}
We were able to measure the fractional linear polarization on all of the four epochs mentioned in paragraph \ref{par:polgeneral}. At 850 MHz, we were not able to measure polarization on 2005 January 10. For the other occasions, the measured polarized fluxes, $P=\sqrt{Q^2+U^2}$, fractions and their error bars are listed in table \ref{tab:linpolfrac}. The latter two quantities are depicted in figure \ref{fig:linpolfracs}. The overall conclusion is that there is no compelling evidence for any significant depolarization at any frequency. Only the polarization fraction at 1300 MHz on January 4 is low compared to the 8.4 GHz measurements, but this fraction was determined from our worst fit, i.e., the fit with the highest reduced $\chi^2$.\\
The polarization angles and their uncertainties are also listed in table \ref{tab:linpolfrac}. The observations at 850 and 1300 MHz gave the most accurate position angles, with typical uncertainties of order $10\degr$. They are depicted in figure \ref{fig:polangles}. Here, we see compelling evidence for significantly different polarization angles with respect to the 8.4 GHz observations from \citet{Taylor2005}, particularly on January 5 and 850 MHz and on January 10 at both 850 and 1300 MHz.

\section{Discussion}
\subsection{Total intensity measurements}
It is clear from figure \ref{fig:fluxPband} that SGR1806-20 is much dimmer at 350 MHz than what would be expected from the GMRT observations at 240 and 610 MHz \citep{Cameron2005}. In principle the Luminous Blue Variable, $14\arcsec$ to the east of SGR 1806-20 \citep[see the Supplementary Information to][]{Gaensler2005a} should be easily distinguishable from the Soft Gamma Repeater in the GMRT images, even at 240 MHz. The FWHM beamsize reported at that frequency is $12\arcsec \times 18\arcsec$ \citep{Chandrab}. This makes it hard to understand the discrepancy.\\
In principle the discrepancy cannot originate from the inclusion or exclusion of extended emission. The GMRT data were corrected for this \citep{Chandraa,Chandrab}. We excluded short spacings ($< 1\mathrm{k}\lambda$) from our 350 MHz WSRT observations. This was actually a necessity since these were daytime observations and solar interference would otherwise compromise our calibration \citep[see also][end of paragraph 3.2]{Brentjens2008}.\\
Also, it is possible that the LBV radio nebula is variable and that it was much brighter on 2005 April 30/May 1 than on some occasions in 2005 January. We ran the AIPS task 'IMFIT' on the map from our 2005 April 30/May 1 observation and we found a peak flux density of $138\pm 1 \mathrm{mJy/beam}$ and an integrated flux of $189\pm 2 \mathrm{mJy/beam}$ at the location of the LBV. The NVSS \citep{Condon98} image of this field shows this source at the 15 mJy level. This would indicate that the LBV has a spectral index of about -1.8, which is almost the index for thermal radio radiation. It should be noted that, at the times of the latest observations in January 2005, when the radio nebula was relatively dim, there is no evidence for negative residuals in our maps that could be caused by the subtraction of the LBV. This indicates that, most likely, the LBV had the same brightness at the times of at least some of the 2005 January measurements as on 2005 April 30/May 1. \\
Variability at radio wavelengths of the radio nebulas from LBVs has been known for quite some time \citep[see, e.g.,][]{Abbott81}. For the P Cygni nebula variability at timescales of days was established at cm wavelengths \citep{Skinner96}. These authors report a $50\%$ increase in flux in less than two days on one occasion during three months of observations on every other day. It is unknown how these variations translate to lower frequencies. We therefore cannot completely exclude that the LBV was brighter at the time of the 2005 April 30/May 1 observation than on some occasions in January 2005. Also, the spectral index derived above does not agree with any of the spectral indices of the four LBVs observed by \citet{Duncan2002} at 3 and 6 cm. Two of those spectral indices are close to that of a spherically symmetric radially expanding stellar wind \citep[+0.6, see][]{Panagia75,Wright75}. However, at these wavelengths, those systems may well be described as optically thin, which may not be the case at the frequencies we are considering.\\
The WSRT 850 MHz Stokes I measurements are not inconsistent with the 840 MHz MOST data published earlier \citep{Gaensler2005a}, given the rather large noise levels in the data from both telescopes. The last MOST observation was taken 15 days after the Giant Flare (GF). Consequently, the 850 MHz WSRT observations after 2005 January 10 cannot be compared with other observations in this band. The last three of the January 2005 observations at 850 MHz were less contaminated by RFI than the first four, which resulted in smaller error bars on the fluxes. There is evidence ($>2\sigma$ level) for a deviation from a power-law decay from about 15 days after the GF, analogous to the 4.8 GHz observations by \citet[][paragraph 2]{Gelfand2005}. These authors also mention a gradual rebrightening from about 25 days after the GF, as a result of swept up ambient material. We can also see that in the WSRT 850 MHz data, but the evidence for this is less compelling, since the sampling of these observations is sparse in time. Consequently, it is shown only in one of our observations, on 2005 January 29, 32.6 days after the GF.

\subsection{Polarimetric measurements}
In figure \ref{fig:linpolfracs} we compare the polarization fractions as listed in table \ref{tab:linpolfrac} with the measurements at 8.4 GHz by \citet[][Table 2]{Taylor2005}. In figure \ref{fig:polangles} we have done the same for the polarization angles. It is clear that the observations at 8.4 GHz are much more accurate. Still, we do not see any significant discrepancies in the polarization fractions. \\
Our observations reveal larger polarization position angles than the 8.4 GHz observations. Most compelling are the observations on 2005 January 5 at 850 MHz and on January 7 at both 850 and 1300 MHz. The error bar on the polarization angle at 350 MHz on January 4 is rather large, but this measurement and the 850 MHz measurement on January 5 show the largest differences with the 8.4 GHz observation, about $85\degr$. At these times, the polarization angles from the 8.4 GHz observations suggest that the magnetic field in the emitting plasma is aligned preferentially along the axis of the radio source, on average \citep{Gaensler2005a}. Thus, the January 4 and 5 polarization angles at 350 and 850 MHz indicate that the magnetic field in the emitting plasma that causes linearly polarized radiation at these low frequencies is close to perpendicular to the axis of the radio source, within $\simeq 20\degr$. Possibly a different substructure in the radio nebula is being probed. It seems hard to explain this feature without a complex model of the radio source. 

\section{Conclusions}
It is striking that depolarization at low frequencies is absent. Also, we have shown that low frequency polarimetry of SGR 1806-20 provides hints with respect to the detailed substructure of the radio nebula which cannot be derived from the extrapolation of high frequency measurements. Models for the radio nebula need to take into account a distinct source of linearly polarized low frequency radiation with magnetic fields in the emitting plasmas aligned quite differently from the fields that are associated with radiation at high frequencies. 

\begin{acknowledgements}
We thank Michiel Brentjens, James Miller-Jones and Gianni Bernardi for helpful discussions about polarization calibration.  We thank Eric Greisen for providing background information about many AIPS tasks. The Westerbork Synthesis Radio Telescope is operated by ASTRON (Netherlands Foundation for Research in Astronomy) with support from the Netherlands Foundation for Scientific Research (NWO). This research was supported by NWO NOVA project 10.3.2.02 (HS) and NWO Vici grant 639.043.302 (BS and RAMJW).   
\end{acknowledgements}

\bibliographystyle{aa}
\bibliography{ref.bib}

\end{document}